\documentclass[
    ,final            
  ]
  {aipproc}

\layoutstyle{6x9}

\input epsf.sty
\input psfig.sty

\def\be{\begin{equation}}
\def\ee{\end{equation}}
\def\lsim{\lower 2pt \hbox{$\, \buildrel {\scriptstyle <}\over
         {\scriptstyle \sim}\,$}}
\def\gsim{\lower 2pt \hbox{$\, \buildrel {\scriptstyle >}\over
         {\scriptstyle \sim}\,$}}

\begin{document}
\title{Emission From \\Rotation-Powered Pulsars and Magnetars}

\author{Alice K. Harding}{
  address={NASA Goddard Space Flight Center, Greenbelt, MD 20771, USA}
}

\begin{abstract}
I will review the latest developments in understanding the high-energy emission of rotation-powered pulsars and magnetically-powered Anomalous X-ray Pulsars (AXPs) and Soft Gamma-Ray Repeaters (SGRs).  These fields have been extremely active in the last few years, both observationally and theoretically, driven partly by new X-ray data from Chandra, XMM-Newton and RXTE.  At the same time, the Parkes Multibeam Survey has discovered over 700 new radio pulsars, some of them young and coincident with EGRET sources, and others having magnetar-strength magnetic fields.  These new observations are raising important questions about neutron star birth and evolution, as well as the properties of their high-energy emission.
\end{abstract}

\maketitle


\section{Introduction}

The last several years have seen significant developments in detection and understanding of 
energetic emission from pulsating neutron stars.   New surveys and detectors have not only discovered more sources 
but have discovered new and unexpected characteristics and behavior of these sources.  Three
types of neutron stars that are the subject of this review, rotation-powered pulsars, Anomalous
X-Ray Pulsars (AXPs) and Soft Gamma-Ray Repeaters (SGRs), were thought to be separate and unrelated
objects as recently as seven years ago.  Today, we know that SGRs and AXPs are both neutron stars 
possessing magnetic fields with unprecedented strength of $10^{14}-10^{15}$ G, collectively called ``magnetars", 
that show both steady X-ray pulsations as well as soft $\gamma$-ray bursts.  Their inferred steady
X-ray luminosities are about one hundred times higher than their spin-down lumnosities, requiring a
source of power beyond the magnetic dipole spin-down that powers radio pulsars.  But some rotation-powered radio 
pulsars have recently been discovered that have magnetic fields, as inferred from dipole spin-down, 
overlapping the range of magnetar fields \cite{Morris02}.  
New high-energy components discovered in the spectra of a number
of AXPs and SGRs require non-thermal particle acceleration and look very similar to high-energy 
spectral components of young rotation-powered pulsars \cite{Kuiper05}.  
So these three types of source class, that were seemly unrelated, are
now more intimately connected in ways that raise interesting questions about their true nature.

At the present time, we treat the radio pulsars and magnetars as separate classes, although they
are both spinning-down neutron stars.  It is believed that all of the observed emission associated 
with radio pulsars is derived from their rotational energy loss or from their heat of formation.  
This includes all pulsed radiation from radio (representing a very small fraction of 
their spin-down luminosity) to $\gamma$-ray wavelengths, as well as surrounding emission from the pulsar 
wind nebula at all wavelengths.  On the other hand, even the average quiescent emission observed from magnetars 
cannot be accounted
for by rotational energy loss, requiring another source of power in SGRs and AXPs,
one that is apparently either not available or not being used by the radio pulsars.  

\begin{figure}
\psfig{figure=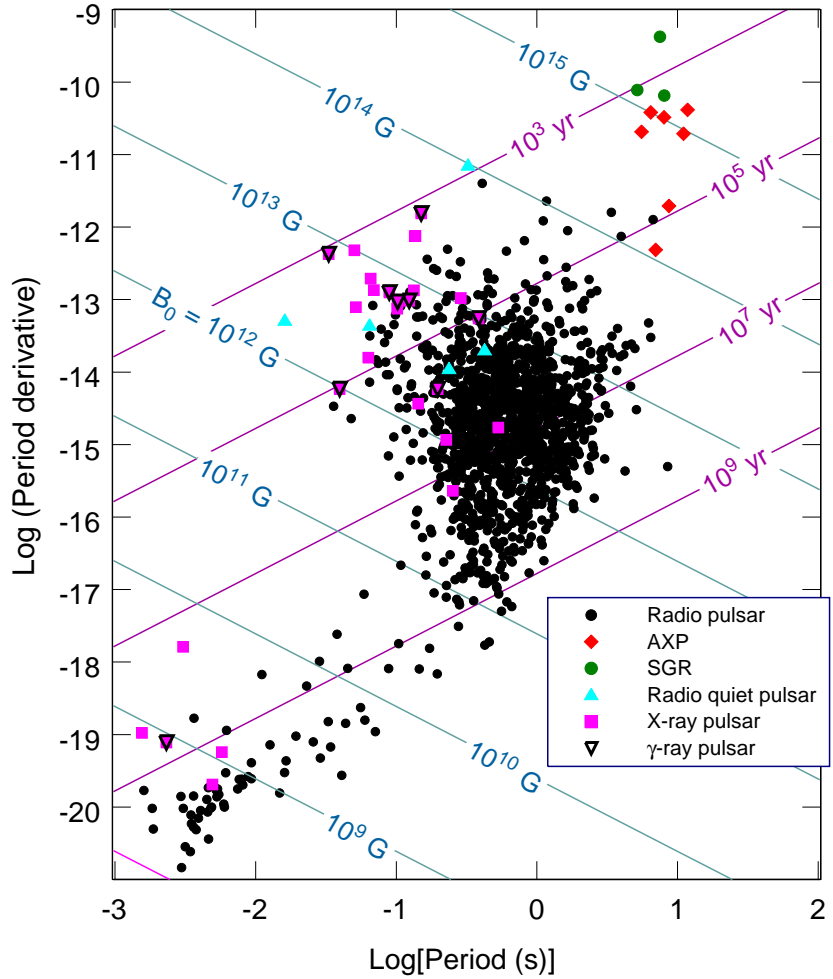,width=11cm,angle=0}
  \caption{Plot of $P$ vs. $\dot P$ for known rotation-powered pulsars and magnetars.  Lines of constant
characteristic age, $P/2\dot P$, and surface dipole field are superposed.}
\end{figure}

\section{Observed Characteristics}

Their source of power is not the only feature that puts rotation-powered pulsars (RPPs) and magnetars in
a separate class.  Most of their observational characteristics are very different, although there
are a few similarities.  Their range of rotation periods, 1.5 ms - 8 s for radio pulsars and 5 - 11 s for 
magnetars, overlap somewhat and they are both observed to spin down more or less steadily with occasional
glitches.  
The measured period derivatives also overlap somewhat.  If plotted together on a $P$-$\dot P$
diagram, their distributions occupy almost, but not quite, separate areas (see Figure 1). 
Both have high energy emission and accelerate particles.  But rotation-powered pulsars and magnetars
display more prominent differences than similarities.  RPPs have never
been observed to have $\gamma$-ray bursts, as is characteristic of SGRs and also recently of AXPs.
Magnetars have very luminous X-ray emission while the RPPs that have magnetar-strength fields have 
very weak or undetectable X-ray emission \cite{wt05}.  
While all but a few known RPPs are radio pulsars, no radio
pulsations have been detected from any SGR or AXP.  The one 
report of radio pulsations from SGR1900+14 \cite{Shitov00} was unconfirmed \cite{lx00} and 
probably confused with emission from the nearby radio pulsar PSR J1907+0918.

A $P$-$\dot P$ plot of known RPPs and magnetars with measured $\dot P$ is shown in Figure 1.
There are presently about 1500 known radio pulsars, but that is expected to change significantly over the
next few years, with several new surveys starting in the Northern hemisphere. 
About 30 radio pulsars are X-ray pulsars and about 9 of these are also $\gamma$-ray pulsars.  The millisecond
pulsars, believed to have been spun-up by accretion torques of a binary companion \cite{Alpar82}, 
have much lower surface magnetic fields in the range of $10^8 - 10^{10}$ G and occupy the lower left part 
of the $P$-$\dot P$ diagram.  There are
a handful of rotation-powered pulsars that were discovered at X-ray or $\gamma$-ray wavelengths and do
not have any detected radio pulsations.
It is not known whether these pulsars are truly radio silent, below sensitivity of radio searches, or
have radio beams pointed in a different direction.  
In the magnetar class, there are 8 known AXPs and 5 SGRs (only 3
of which have measured period derivatives).  

\subsection{Rotation-powered pulsars}

Our knowledge of high-energy emission from RPPs has increased dramatically over the past ten years.
The Compton Gamma-Ray Observatory (CGRO) discovered five new $\gamma$-ray pulsars which, added to the 
Crab and Vela $\gamma$-ray pulsars that were previously known, brings the number to seven that are known
with high significance \cite{Kan02}.  
There are about five others that were detected with less significance, one of
these being the ms pulsar PSR 0218+4232, which is also an X-ray pulsar.  Although the number of $\gamma$-ray
pulsars is still small, some interesting patterns in their emission characteristics have emerged.
Their luminosity (really their flux times distance squared, since the solid angle is not known) follows the
relation \cite{Thom04}, $L_{HE} =  {L_{\rm SD}}^{1/2}$, where $L_{\rm SD}$ is the spin-down luminosity.  
Equivalently, the high energy luminosity is
proportional to the voltage across the open field lines, or $L_{HE} \propto V_{PC} = 
4 \times 10^{20}\,P^{-3/2}\dot P^{1/2}$ V
and to the Goldreich-Julian flux, the maximum particle flux emerging over the polar caps. 
Their spectral index above 100 MeV shows a decrease with characteristic age of the pulsar, with photon spectral
indices ranging from -2.2, in the case of the Crab, to -1.2 in the case of PSR B1055-52. All $\gamma$-ray
pulsar spectra show evidence for a high energy turnover, either in the EGRET band around a few GeV or implied
sharp breaks or turnovers between 10 and 100 GeV from upper limits above 100 GeV, from searches by Air Cherenkov Telescopes \cite{Thom04}.  No pulsed emission has been
detected above 20 GeV in any pulsar.   The $\gamma$-ray
light curves generally show two narrow peaks separated by 0.3 - 0.5 in phase, so that the profiles are very
broad compared to radio pulse profiles, that usually occupy a small fraction of phase.  But in the case of 
the two shortest period $\gamma$-ray pulsars, the Crab and ms PSR J0218+4232, the radio profile is as broad as the
$\gamma$-ray profile and the peaks are correlated in phase.  For the other $\gamma$-ray pulsars, single narrow
radio peaks lead the double $\gamma$-ray pulses in phase.  This behavior possibly suggests a different
relation between radio and high-energy emission sites in short period and longer period pulsars.

A much larger number of pulsars are visible in the X-ray band \cite{krh05}, and the population 
of X-ray pulsars is
more diverse, containing many more middle-aged and older sources.  The spectra often show both non-thermal 
or power law components (probably magnetospheric in origin) and one or more thermal components (originating from cooling of the whole neutron star surface and/or heated polar caps).   The non-thermal spectrum is harder in the 
X-ray band than in the $\gamma$-ray band, so that the spectrum breaks in the soft $\gamma$-ray band.
In the young, Crab-like pulsars the magnetospheric emission significantly dominates over any surface thermal emission. In the middle-aged pulsars multiple thermal components are often present, a cool blackbody with $T \sim 
0.5 - 1$ MK and a hot blackbody with $T > 1$ MK.  As the neutron star ages, the cooling component decreases rapidly, so that in old pulsars (including the millisecond pulsars) the heating component may dominate. 
The luminosity in the X-ray band has a different dependence on spin-down luminosity \cite{BT97}, 
$L_X \simeq 10^{-3}\,L_{\rm SD}$, although there is a very large scatter in the correlation.  
Since there are many more 
measured pulse profiles of X-ray pulsars, there is a larger data base to compare radio and high-energy profiles.
The trend of greater multiwavelength phase coherence in the profiles of fast pulsars, that we are beginning 
to see for $\gamma$-ray pulsars, seems also true at X-ray wavelengths.  PSR B0540-69, a 50 ms pulsar in the LMC,
and PSR J1617-5055, a 69 ms pulsar in RCW103, show X-ray and radio pulses that are very similar and in phase.
There is also phase coherence of X-ray and radio pulses in ms pulsars PSR J0218+4232, PSR B1821-24, 
PSR J0437-4715 and PSR B1937+21.  

The broad-band spectra of pulsars show several trends \cite{Thom04}.  The emission power peaks in hard X-rays
for younger pulsars like the Crab and PSR B1509-58, so that thermal components are not visible.  For older pulsars the power peaks around several GeV and the non-thermal power in X-rays is much smaller so that the thermal
components become quite prominent.   Quite a few have optical pulsations whose spectra often connect to the
non-thermal X-ray components.  The spectra of ms pulsars behave differently from either of these groups, with
very hard X-ray spectra (photon indices of -0.6 -- -1.2) and weak detections or upper limits above 100 KeV that
imply a power peak around 1 MeV \cite{kh03}.

Although EGRET significantly detected $\gamma$-ray pulsations from only seven pulsars, it detected around 150
point sources that have no identified counterparts \cite{Hart99}.  
Many of these lie near the Galactic plane and may be
pulsars.  Since the CGRO mission ended in 2000, there have been over 700 new pulsars discovered \cite{Man01}, 
and around 50
lie inside EGRET source error circles \cite{Kramer03} \cite{Grenier04} .  About 20 of these are plausible
counterparts both on energetic grounds or from model estimates \cite{hm05}. 
These sources unfortunately contain
too few photons to perform any credible period searches of the EGRET archival data with present pulsar 
ephemerides, so identification must await more $\gamma$-ray observations. 

\subsection{Magnetars}

SGRs and AXPs were both discovered several decades ago, but they have only recently been recognized as similar objects (for detailed review, see Woods \& Thompson \cite{wt05}).  
SGRs were first detected around 1979 as $\gamma$-ray transients and were thought to be a type of 
classical $\gamma$-ray burst.  They undergo repeated bursts with several tenths of second duration and average
energy $10^{40}-10^{41}\,\rm erg$, and their bursting often occurs in episodes spaced years apart.  
They more rarely undergo giant superflares of total energy $10^{45}-10^{47}\,\rm erg$, consisting of an initial
spike of duration several hundred ms followed by a longer pulsating decay phase of duration several hundred seconds.  Such superflares have been observed in three SGR sources, SGR0526-66 (the famous 5th March 
1979 event), SGR1900+14 \cite{Hurley99} and very recently in SGR1806-20 \cite{Palmer05}.  
In 1998, SGR1806-20 was discovered to have 7.47 s pulsations in its quiescent X-ray emission \cite{Kouv98}
and a large $\dot P$ that implies a surface magnetic field of $\simeq 10^{15}$ G
if due to dipole spin-down.  Quiescent periodicities of 8 s and 5.16 s and large $\dot P$ were subsequently 
detected in SGR0526-66 and SGR1900+14, implying similarly high surface magnetic fields.  In all three sources, 
the quiescent periods are the same periods seen in the decay phases of their superflares.  The quiescent 
pulse profiles are very broad and undergo dramatic changes just before and after superflares.  The profiles
are often more complex, with multiple peaks before flares, changing to more simple single peaks profiles
following the flares.  All of the SGRs
lie near the Galactic plane and are thought to have distances around 10-15 kpc (except for SGR0526-66, which is
in the LMC).

The first AXPS were discovered as pulsating X-ray sources in the early 1980s by EXOSAT and were thought
to be a strange (anomalous) type of accreting X-ray pulsar.  They are
bright X-ray sources possessing luminosities (in their highest states) of $L_X \sim 10^{35}\,\rm
erg\; s^{-1}$, but show no sign of any companion or accretion disks that would be 
required to support the accretion hypothesis.  The AXPs have pulsation periods
in a relatively narrow range of 5 - 11 s and are observed to be spinning down with large period derivatives \cite{vg97}.  Their pulse profiles are broad and very similar to those of SGR sources.
The very high surface magnetic fields of $10^{14}-10^{15}$ G implied 
by dipole spin-down were originally controversial, but have come to be accepted after the quiescent
periods were found in SGRs and especially following the recent discovery of SGR-like bursts from several AXPs \cite{Kaspi03}.  It is now believed that SGRs and AXPs are two varieties of the same type of object, 
very strongly magnetized, isolated neutron stars possibly powered by magnetic field decay.  In both sources,
the high-state quiescent luminosities of $L_X \sim 10^{35}\,\rm erg\; s^{-1}$ 
are much higher than their spin-down
luminosities of $L_{\rm SD} \sim 2-6 \times 10^{33}\,\rm erg\; s^{-1}$, 
demanding an alternative power source.

Periodic monitoring of AXPs have revealed a significant amount of flux variability in the
sources.  The luminosity of several AXPs has even changed by several orders of magnitude.
1E 1048.1-5937 has been observed to undergo pulsed flux flares lasting a few weeks \cite{gk04} , and the baseline luminosity between such flaring events is only 
$6 \times 10^{33}\,\rm erg\; s^{-1}$ 
which is close to the spin-down luminosity.  Since the sources were discovered in their high
states, such extended low states were not observed until recently.

The quiescent spectra of AXPs and SGRs consist of a thermal component fit by $\sim$ 0.5-1 keV blackbodies
and one or more non-thermal components.  Until recently, the non-thermal spectra below 10 keV were fit with 
steep power laws having indices $\sim -3$ -- $-4$.  When {\sl INTEGRAL} and {\sl RXTE} recently 
measured the spectra above 10 keV for the first time hard, non-thermal components were discovered in
three AXPs, and also SGR 1806-20.  In two of the AXPs, the differential
spectra between 10 keV and 50 keV are extremely flat: 1E 1841-045 \cite{khm04}
 has a power-law index of $s = -0.94$ and
4U 0142+61 \cite{Hartog04} displays an index of $s = -0.45$,
both much flatter than the steep non-thermal components in the classic
X-ray band.  RXS J1708-40 possesses a slightly steeper continuum with
$s = -1.18$.  The non-thermal tail of quiescent emission in SGR 1806-20 is
similarly pronounced \cite{Mereg05} \cite{Molkov05}, but
somewhat steeper, with an index of $s= -1.6$ -- $-1.9$ extending to 100 keV.  Such hard non-thermal
components require continuous particle acceleration during the quiescent state.

\section{Emission Models}

\subsection{Spin Power}

The characteristics of emission from rotation-powered pulsars have been studied in great detail 
for many years.  Even so, the origin of the pulsations is still not understood at any wavelength 
except to realize that the radio emission must be a coherent process
requiring significant particle densities and probably electron-positron pairs  \cite{Melrose00}, and
the emission at higher energies comes from incoherent processes requiring acceleration of particles
to energies of at least 10 TeV.  The mechanisms and models for pulsed high-energy emission from 
rotation-powered pulsars
was traditionally based on the physics of particle acceleration, which results from the strong
electric fields ($E_{\parallel}$) that are induced along open magnetic field lines.  Two accelerator 
sites have been studied in most detail and two main types of model have developed, although pulsed emission
from the wind zone has also been proposed \cite{pk05}.  Polar cap
models \cite{dh96} are based on particle acceleration near the magnetic poles and outer gap models \cite{CHR86, 
Rom96} are based on
acceleration in vacuum gaps in the outer magnetosphere, however the emission geometries of both models 
have significantly evolved in recent years .

\subsubsection{Classic polar cap model}

Polar cap models divide into two types that depend on how the Goldreich-Julian 
(or corotation) charge density, which would cause force-free conditions, is supplied and distributed.  
Vacuum gap accelerators \cite{rs75} assume that charges are 
trapped in the neutron star surface so that the full vacuum potential drop can develop above the surface.
Such accelerators are unstable to break-down by electron positron pair creation and are therefore non-steady accelerators.  Space charge-limited flow (SCLF) accelerators \cite{saf78}
assume that charge is freely supplied by the neutron star, so that the charge density at the stellar surface is equal to the corotation charge density.  
Although $E_{\parallel} = 0$ at the neutron-star surface in these models, the space charge along open 
field lines above the surface deviates from the corotation charge, due to the curvature of the field 
 \cite{as79} and to general relativistic inertial frame dragging \cite{MT92}, causing a steady $E_{\parallel}$
that increases with height.  Accelerated particles radiate $\gamma$-rays that create electron-positron pairs
and the $E_{\parallel}$ is screened above a pair formation front (PFF) by polarization of the pairs.  
The potential drop is thus self-adjusted to give particle Lorentz factors around $10^7$, 
assuming a dipole magnetic field.
Above the PFF, force-free conditions could develop if the pair multiplicity is sufficient.  This is likely
for relatively young pulsars, which can produce pairs through curvature radiation \cite{HM01}, 
but older pulsars that can
produce pairs only through inverse Compton emission are expected to be pair starved and their open magnetospheres
would not achieve a force-free state \cite{HM02,MH04b}.
Given the small radiation loss length scales for particles of these energies, the high energy radiation will
occur within several stellar radii of the surface.  One thus expects that the radiation spectra will exhibit 
sharp high-energy cutoffs due to magnetic pair production attenuation at energies in the range 0.01-20 GeV,
depending inversely on field strength.
The radiation from electromagnetic cascades produces a hollow cone of emission 
around the magnetic pole, with opening angle determined by the polar cap half-angle, 
$\theta_{_{PC}}$, at the radius of emission $r$.  The characteristics of emission from this
type of polar cap model \cite{dh96}  have been successful in reproducing many 
features of $\gamma$-ray pulsars, including the wide double-peaked pulses observed in $\gamma$-ray pulsars 
like the Crab, Vela and Geminga, and the phase-revolved spectra.  
However, since the polar cap opening angle is very small (a few degrees), 
the broad profiles which require beam opening 
angles of the order of the magnetic inclination angle, cannot be produced 
unless the pulsar is nearly aligned.

More recent versions of the polar cap model \cite{MH03, MH04a} have explored acceleration in the
`slot gap', a narrow region bordering the last open field line in which the electric field is unscreened. 
Near the open field line boundary, the $E_{\parallel}$ is lower and a larger distance is required for the electrons to accelerate to the Lorentz factor needed to radiate photons energetic enough to produce pairs.  The PFF thus occurs at higher and higher altitudes and curves upward as the boundary is approached, approaching infinity and becoming asymptotically parallel to the last open field line.  If the $E_{\parallel}$ is effectively screened above the PFF, then a narrow slot surrounded by two conducting walls is formed.   Pair cascades therefore do not take place near the neutron star surface in the slot gap, as do the pair cascades along field lines closer to the magnetic pole, but occur on the inner edge of the slot gap at much higher altitudes.  The high energy emission beam is a
hollow cone with much larger opening angle than that of the lower altitude cascade emission.

\subsubsection{Caustics, extended slot gaps and outer gaps}

Morini \cite{Morini83} first noted an interesting feature of the geometry of emission in the outer magnetosphere 
of a rotating dipole.  If one assumes that photons are radiated tangent to the magnetic
field from the polar cap to the light cylinder, 
then the relative phase shifts of photons emitted at 
different radii due to dipole curvature, aberration and time-of-flight nearly cancel on  
field lines on the trailing edge of the open region.  Radiation along such trailing field lines 
bunches in phase, forming a sharp peak in 
the profile.  On the other hand, photons emitted at different radii along leading field lines spreads 
out in phase.  The effect is most pronounced for large inclination of the magnetic axis to the rotation axis.
A plot of observer angle to the rotation axis versus phase, as shown in Figure 2a and 2b, clearly displays the
sharp lines of emission, or caustics, along the trailing field lines originating at radii 
$r_{\rm em} \sim 0.2 - 0.8\,R_{LC}$, where $R_{LC} = c/\Omega$ is the light cylinder radius.
Morini's model for the Vela pulsar \cite{Morini83} considered the caustic emission from only one magnetic pole.  
Dyks \& Rudak \cite{dr03} explored a purely geometrical model in which they assume emission along the last open
field lines of both magnetic poles and discovered that such a two-pole caustic model (see Figure 2) can naturally reproduce 
the main features of $\gamma$-ray pulsar profiles.  

Several different types of pulsar emission models, including the extended slot gap and outer gap, 
feature caustics.  The extended slot gap model is a consequence of the SCLF polar cap models and 
originates from the fact that the 
potential in the slot gap is unscreened, so that electrons on field lines which thread the slot 
gap will continue accelerating to very high altitudes \cite{MH04a}, maintaining Lorentz factors $\sim 10^7$.
The emission pattern exhibits both the hollow cone centered on the magnetic poles, from the low-altitude
cascades, as well as caustics from high-altitude emission in the slot gap along the trailing field lines. 
Thus, the slot gap may provide a physical basis for the two-pole caustic model, although a full radiation
model of the slot gap remains to be developed.

\begin{figure} 
\epsfysize=14cm
\epsfbox{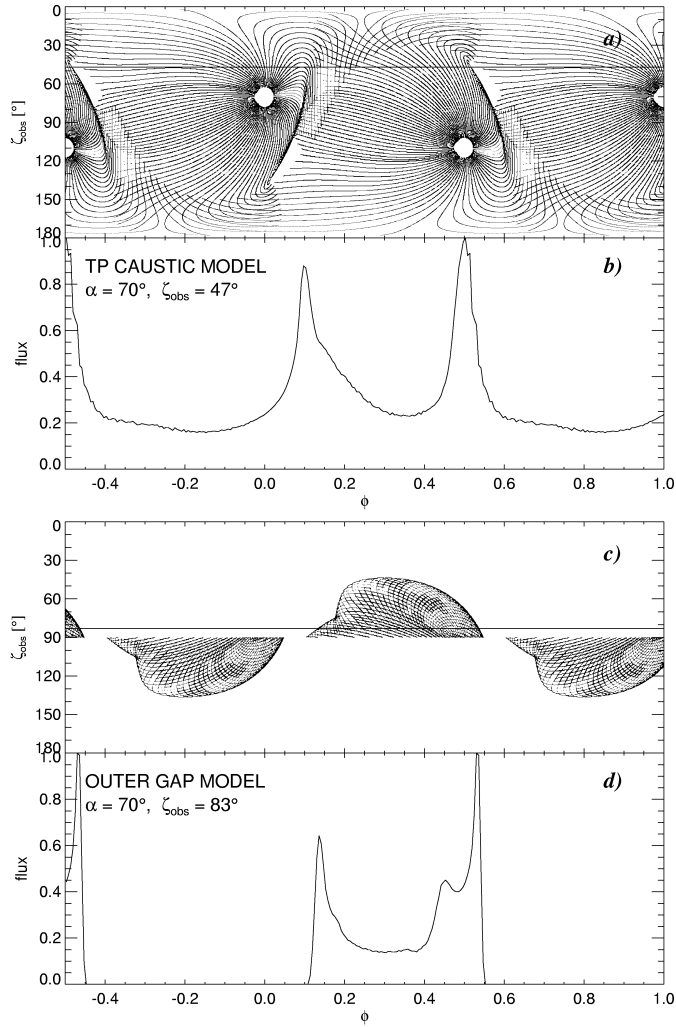}

\caption{Comparison of two-pole caustic and outer gap models\cite{DHR04}. 
a) Plot of emission in the ($\zeta_{\rm obs}$, $\phi$) plane, calculated with a two-pole caustic
model for inclination, $\alpha = 70^{\circ}$.  $\phi$ is the rotational phase and $\zeta_{\rm obs}$ is the 
viewing angle measured from the rotation axis. b) High-energy profile for $\zeta_{\rm obs} = 47^{\circ}$, 
produced by a horizontal cut through the phase plot above.  c) As in a) for an outer gap model.
d) High-energy profile from the outer gap phase plot c) for $\zeta_{\rm obs} = 83^{\circ}$}
\end{figure}

Outer-gap models   \cite{CHR86,Rom96} assume that acceleration occurs in vacuum gaps that 
develop in the outer magnetosphere, along the last open field line above the null charge surfaces, 
where the Goldreich-Julian charge density changes sign, and that 
high-energy emission results from photon-photon pair production-induced cascades.  The pair
cascades screen the accelerating electric field and limit the size of the gap both along and 
across the magnetic field.  The geometry
of the outer gaps prevents an observer from seeing gaps associated with both magnetic poles, putting
them in the class of one-pole caustic models.  The high energy pulse profiles that are observed to 
have two widely separated, sharp peaks are formed by caustics, the leading peak originating from 
overlapping field lines at $r_{\rm em} \sim 0.9\,R_{LC}$ and the trailing peak originating from the 
caustic along trailing field lines at $r_{\rm em} \sim 0.2 - 0.8\,R_{LC}$.  An example of outer gap 
profile formation is shown in Figure 2c and 2d.  
Some recent outer gap models point out that in general the gap lower boundary should exist
somewhat below the null charge surface, and the location of this boundary will depend on the external 
currents that flow into or out of the gap  \cite{HSH03, TSH04}.  If the gap moves close enough to the
neutron star, depending on inclination angle, it might be possible for an observer to see gaps from both
magnetic poles.

Although outer gap and slot gap models can produce similar pulse profiles, the phase plots of their emission 
show prominent differences.  As displayed in Figure 2a, slot gap/two-pole caustic emission fills the entire sky
and the emission can be viewed at a wide range of angles.  Radiation
will be visible at all pulse phases so that pulse profiles like the one shown in Figure 2b will
include `off-pulse' emission. 
Outer gap emission fills only a fraction of the sky, as shown in Figure 2c, since outward-going radiation
occurs only above the null charge surface in each hemisphere \cite{crz00}.  
Outer gap emission is therefore restricted to large inclination and viewing angles $\alpha, \zeta_{\rm obs} 
\gsim 50^{\circ}$.  The peaks in outer gap 
profiles drop sharply at their outer edges and there is no off-pulse emission outside the peaks, 
as is evident in Figure 2d.

\subsection{Magnetic Power}

Magnetars were actually predicted by Duncan \& Thompson (DT) \cite{dt92} before their existence was 
observationally verified.
In this model, some neutron stars born with ms periods $P_i$ generate huge magnetic fields of 
$3 \times 10^{17}{\rm G}\,(P_i/1 \rm ms)^{-1}$ by 
dynamo action soon after their birth.  Such high fields have several properties that could distinguish
behavior of magnetars from that of neutron stars having lower fields.  Magnetar fields can decay on much 
shorter timescales,
\be
t_{amb}^{}  \cong 10^5 yr_{} \left( {\frac{{B_{core} }}{{10^{15} G}}} \right)^{ - 2} 
\ee
due to ambipolar diffusion.
Diffusion of magnetic flux out of the neutron star core on these timescales provides the power to magnetars in the 
DT model.  Magnetar-strength fields also apply higher stresses to the stellar crust, so that the yield
strain can exceed the crustal strength.  This property is responsible for the small SGR and AXP bursts in
the DT model \cite{td96}.  If a toroidal component of the field $B_{core} > 10^{15}$ G develops in the interior of the star, it can twist the external field \cite{dt01}.  
Such action can cause the superflares if the twisted field lines reconnect.
Finally, due to the much faster heat transport in very strong magnetic fields, there is a greater heat flux
through the crust \cite{hk98}.  Such a property may explain the much hotter surface 
temperatures of magnetars and the high quiescent X-ray emission.

Magnetar fields also induce significant changes in radiative processes (see  \cite{Duncan00} and  \cite{Harding02} for reviews).  In general, in magnetic fields approaching and exceeding the quantum critical
field $B_{\rm cr} = 4.4 \times 10^{13}$ G, radiative processes such as Compton scattering, cyclotron an 
synchrotron emission and absorption, and pair production and annihilation, classical descriptions are
largely inaccurate and QED descriptions must be used.  In addition, new processes become possible in strong
fields, such as one-photon pair production and annihilation, vacuum polarization and photon splitting, 
that cannot take place in field-free environments.  These processes, in particular vacuum polarization \cite{Lai05}, strongly influence the propagation of photons in the neutron star atmosphere and the spectrum of the 
emergent radiation.  

According to the DT model \cite{td95}, the magnetar superflares result from reconnection of sheared or twisted external
field lines, leading to particle acceleration and radiation of hard emission.  The estimated luminosity of
such events,
\be
E_{\rm Flare} \approx \frac{{B_{core}^2 }}{{8\pi }}R^3  \approx 4 \times 10^{46} {\rm erg} \left( {\frac{{B_{core} }}{{10^{15} G}}} \right)^2, 
\ee
is similar to observed luminosities of superflares.  The smaller bursts result from cracking of the crust,
which is continually overstressed by diffusion of magnetic flux from the neutron star interior.  The shaking of 
magnetic footpoints then excites Alfven waves that accelerate particles.  The energy radiated in such
events would be
\be
E_{SGR}  \cong 10^{41} {\rm erg} \left( {\frac{{B_0 }}{{10^{15} G}}} \right)^{ - 2} _{} \left( {\frac{l}{{\rm 1 km}}} \right)^2 _{} \left( {\frac{{\theta _{\max } }}{{10^{ - 3} }}} \right)^2,
\ee
where $l$ is the length scale of the displacements, $B_0$ is the crustal field and 
$\theta _{\max }$ is the yield strain of the crust.
The quiescent emission in the DT model is powered by magnetic field decay through conduction of heat from
the core.  The neutron star crust is heated to a temperature of 
$T_{crust}  \cong 1.3 \times 10^6 K (T_{core}/10^8 K)^{5/9}$ 
where $T_{core} $ is the core temperature, providing a luminosity
\be
L_{\rm{x}}  \cong 6 \times 10^{35} {\rm erg s^{-1}} \left( {\frac{{B_{core} }}{{10^{16} G}}} \right)^{4.4}.
\ee
Thompson \& Beloborodov \cite{tb04} have proposed that the hard, non-thermal quiescent component is due to the
creation of a strong $E_{\parallel}$ induced by twisting of field in the closed region, producing synchrotron radiation from electron acceleration at high altitude. 

An alternative model for magnetar activity and emission has been discussed by 
Heyl \& Hernquist \cite{hh05a}. 
The burst and quiescent radiation are a result of shocks from fast-mode plasma waves.  The hard quiescent 
component is due to a pair-synchrotron cascade \cite{hh05b}.  Baring \cite{Baring04} proposes that 
resonant Compton upscattering 
of thermal X-rays by accelerated particles in the open field region produces the quiescent hard  emission.        

\section{Unresolved Questions and Future Answers}

Despite all the progress made in understanding the nature and working of rotation-powered
pulsars, SGRs and AXPs over the last decade, there is still a wealth of questions remaining to be answered.
Among the most vexing is the unknown fundamental difference that distinguishes rotation-powered 
pulsars from magnetars.  
Among the most persistent and enduring unresolved issues is the geometry of pulsar accelerators.
In addition, we would like to know how magnetars accelerate particles and if AXPs have giant flares like SGRs.
Current observations by INTEGRAL are beginning to uncover clues to magnetar particle acceleration.  It is 
hoped that SWIFT may find the nature of the short $\gamma$-ray bursts and their possible relation to SGR bursts. 
Future high-energy $\gamma$-ray telescopes such as AGILE and GLAST will detect many more pulsars, possibly
several hundred in the case of GLAST, many of them radio quiet.  With this large increase in statistics, as well
as more detailed phase-resolved spectroscopy, we may be able to map out the geometry (and possibly even uncover the physics) of pulsar particle acceleration.


  





\end{document}